\documentclass[conference]{IEEEtran}
\IEEEoverridecommandlockouts

\def\BibTeX{{\rm B\kern-.05em{\sc i\kern-.025em b}\kern-.08em
    T\kern-.1667em\lower.7ex\hbox{E}\kern-.125emX}}

\usepackage{tikz}
\usepackage[caption=false,font=footnotesize]{subfig}
\usepackage{graphics}
\usepackage{algorithm,algpseudocode}
\usepackage{float}
\usepackage{multirow}
\usepackage[outdir=./]{epstopdf}
\usepackage{amsmath}
\usepackage{caption}
\usepackage{multicol}
\algnewcommand{\LineComment}[1]{\State \(\triangleright\) #1}
\begin{document}
\DeclareRobustCommand*{\IEEEauthorrefmark}[1]{%
	\raisebox{0pt}[0pt][0pt]{\textsuperscript{\footnotesize\ensuremath{#1}}}}
\title{Hybrid Compression Techniques for EEG Data Based on Lossy/Lossless Compression Algorithms}
\author{\IEEEauthorblockN{ Madyan Alsenwi\IEEEauthorrefmark{1},
		Tawfik Ismail\IEEEauthorrefmark{2}, and
		M. Saeed Darweesh\IEEEauthorrefmark{3,}\IEEEauthorrefmark{4}}
	\IEEEauthorblockA{\IEEEauthorrefmark{1}Department of Computer Science and Engineering, Kyung Hee University, Gyeonggi-do 17104, South Korea}
	\IEEEauthorblockA{\IEEEauthorrefmark{2} National Institute of Laser Enhanced Science, Cairo University, Giza, Egypt}
	\IEEEauthorblockA{\IEEEauthorrefmark{3} Institute of Aviation Engineering and Technology, Giza, Egypt}
	\IEEEauthorblockA{\IEEEauthorrefmark{4} Nanotechnology Department, Zewail City for Science and Technology, Egypt}
	Email: malsenwi@khu.ac.kr; tismail@niles.cu.edu.eg;  msaeed@ieee.org;}
\maketitle
\begin{abstract}
The recorded Electroencephalography (EEG) data comes with a large size due to the high sampling rate. Therefore, large space and more bandwidth are required for storing and transmitting the EEG data. Thus, preprocessing and compressing the EEG data is a very important part in order to transmit and store it efficiently with fewer bandwidth and less space. The objective of this paper is to develop an efficient system for EEG data compression. In this system, the recorded EEG data are firstly preprocessed in the preprocessing unit. Standardization and segmentation of EEG data are done in this unit. Then, the resulting EEG data are passed to the compression unite. The compression unit composes of a lossy compression algorithm followed by a lossless compression algorithm. The lossy compression algorithm transforms the randomness EEG data into data with high redundancy. Subsequently, A lossless compression algorithm is added to investigate the high redundancy of the resulting data to get high Compression Ratio (CR) without any additional loss. In this paper, the Discrete Cosine Transform (DCT) and Discrete Wavelet Transform (DWT) are proposed as a lossy compression algorithms. Furthermore, Arithmetic Encoding and Run Length Encoding (RLE) are proposed as a lossless compression algorithms. We calculate the total compression and reconstruction time (T), Root Mean Square Error (RMSE), and CR in order to evaluate the proposed system. Simulation results show that adding RLE after the DCT algorithm gives the best performance in terms of compression ratio and complexity. Using the DCT as a lossy compression algorithm followed by the RLE as a lossless compression algorithm gives $CR=90\%$ at $RMSE=0.14$ and more than $95\%$ of CR at $RMSE=0.2$. 
\end{abstract}
\section{Introduction}
Electroencephalography (EEG) is an electrophysiological monitoring mechanism used to record and evaluate the electrical activity in the brain. A number of sensors are attached to the scalp or placed inside the human body called electrodes. These electrodes record the electrical impulses in the brain and send it to an external station for analyzing it and recording the result. The recorded EEG data have a large size due to the high sampling frequency and transmitting these large size of data is difficult due to the channel capacity and power limitations.  

Data compression is proposed as a solution to handle the problem of channel capacity and power consumption limitation. Generally, The compression methods are classified as lossless compression techniques and lossy compression techniques. The original data in the lossless compression algorithms can be reconstructed perfectly without any distortion in the original data. Conversely, the lossy compression techniques lead to non perfect reconstruction since some parts of the original data may be loosed. However, Higher Compression Ratio (CR) can be achieved using the lossy compression algorithms compared with the lossless compression techniques \cite{Alsenwi2017}.    

The randomness nature of the EEG data makes it difficult to achieve a high CR using only lossless compression techniques \cite{birvinskas2015fast,Alsenwi2016}. Thus, lossy compression algorithms with accepted level of distortion are used in this work. Firstly, the original EEG data are standardized, to give it the property of standard normal distribution, and segmented in the preprocessing unit. Then, a lossy compression algorithm followed by lossless compression algorithm are applied to the preprocessed data. In this paper, the Discrete Cosine Transform (DCT) and Discrete Wavelet Transform (DWT) are used as a lossy compression algorithms. The output data from the lossy compression algorithm (DCT/DWT) have high redundancy. Therefore, adding a lossless compression algorithm after the lossy compression algorithm gives a high CR without any additional loss in the EEG data. Run Length Encoding (RLE) and Arithmetic Encoding (Arithm) are used as a lossless compression algorithms in this work. 

Several works have been studied the compression of EEG data \cite{hadjileontiadis2006biosignals}. Authors in \cite{hadjileontiadis2006biosignals} used only the DCT algorithm, which is a lossy compression technique, for EEG data compression. using only lossy compression cannot give a high CR compared with the case of using lossy followed by lossless compression algorithm. Authors in \cite{akhter2010ecg} proposed a compression system composed from DCT and RLE followed by Huffman Encoding. This compression system can give high CR but it is more complex and long time is required for compression and reconstruction processes. Authors in \cite{deshlahra2013comparative} introduced a comparative analysis between DWT, DCT, and Hybrid (DWT+DCT). A high distortion can be occurred in the EEG data due to using lossy compression algorithm (DCT) followed by another lossy compression algorithm (DWT). In this work, we propose a lossy compression algorithm followed be a lossless compression algorithm in order to balance between the data loss, CR, and the system complexity. Different combinations of lossy and lossless compression algorithms are studied in order to achieve the combination that gives the best performance.    

The rest parts of this paper are organized as follows: Section II introduces an overview on the compression techniques. Both DWT, DCT, Arithm, and RLE are described in this section. Section III presents the proposed compression model and the performance metrics. Simulation results are discussed in section IV. Finally, section V concludes the paper.  

\section{Data Compression Techniques}
An overview of the data compression algorithms used in this paper is presented in this section. Generally, the data compression techniques are classified as lossy and lossless compression techniques \cite{antoniol1997eeg,koyrakh2008data}. A brief description of the lossy/lossless compression algorithms used in this work is given bellow:
\subsection{Discrete Cosine Transform (DCT)}
Discrete Cosine Transform (DCT) is a transformation technique that transforms a time series signal into its frequency components. The main feature of DCT is its ability to concentrate the input signal energy at the first few coefficients of the output signal. This feature is investigated widely in the data compression field.

Let $f(x)$ be the input EEG signal of the DCT which composes of $N$ EEG data samples and let $Y(u)$ be the output signal of DCT which composes of $N$ coefficients. The one dimensional DCT is given by the following equation \cite{salomon2004data,fauvel2014energy}:
\begin{equation}
Y(u)=\sqrt{\frac{2}{N}}\alpha(u)\sum_{x=0}^{N-1}f(x)\:cos(\frac{\pi(2x+1)u}{2N})
\label{DCT}
\end{equation}
where\\
\[\alpha(u) = \left\{
\begin{array}{lr}
\frac{1}{\sqrt{2}}, & u=0\\
1, & u>0
\end{array}
\right.\]
The first coefficient of $Y(u)$, $Y(0)$, is the DC component and the remaining coefficients are referred as AC components. The DC component $Y(0)$ is the mean value of the original signal $f(x)$ whereas the AC components represent the frequency of the $f(x)$ and it is independent on the average. The inverse process of DCT takes the coefficients of $Y(u)$ as an input and transforms it back into $f(x)$. The inverse transform of DCT is given as follows:
\begin{equation}
f(x)=\sqrt{\frac{2}{N}}\alpha(u)\sum_{u=0}^{N-1}Y(u)\: cos(\frac{\pi(2x+1)u}{2N})
\label{IDCT}
\end{equation}
Most of the coefficients produced by DCT have a small values and usually approximated to zero. 
\subsection{Discrete Wavelet Transform (DWT)}
\begin{figure}
	\centering
	\includegraphics[width=\linewidth]{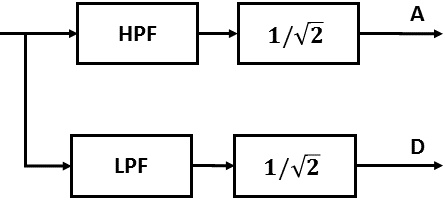}
	\captionof{figure}{DWT Tree}
	\label{DWT_tree}
\end{figure}

Discrete Wavelet Transform (DWT) decomposes the input signal into high frequency part called details and low frequency part called approximation as shown in Fig. \ref{DWT_tree}. This decomposition of the input signal allows studying each frequency component with a resolution matched to its scale and investigated in the data compression \cite{rajoub2002efficient,shaeri2015method}. 
DWT with Haar basis function is used in this study because it is less complex and gives good performance. Every two consecutive samples ($S(2m), S(2m+1)$) in DWT with Haar functions has two coefficients defined as \cite{deshlahra2013comparative,khalifa2008compression}:
\begin{equation}
C_{A}(m)=\frac{1}{\sqrt{2}}[S(2m)+S(2m+1)]
\label{Approximation coeff}
\end{equation}
\begin{equation}
C_{D}(m)=\frac{1}{\sqrt{2}}[S(2m)-S(2m+1)]
\label{details coeff}
\end{equation}
Where $C_A(m)$ is the low frequency component (approximation coefficient) and $C_D(m)$ is the high frequency component (detail coefficient). Equations 3 and 5 show that calculating the $C_A(m)$ and $C_D(m)$ is equivalent to pass a signal through high-pass and low-pass filters with subsampling factor of 2 and normalized it by $1/\sqrt(2)$.
\subsection{Run Length Encoding (RLE)}
\begin{figure}
	\centering
	\includegraphics[width=\linewidth]{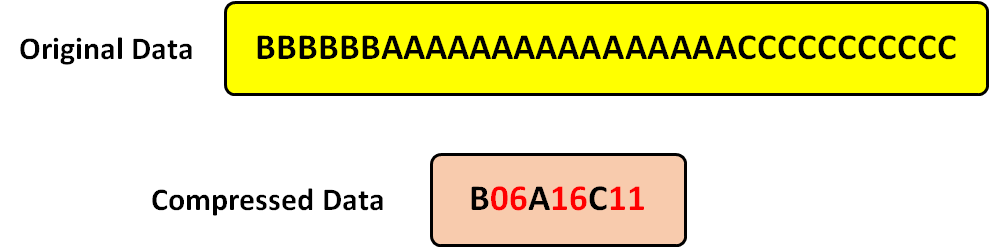}
	\captionof{figure}{RLE} \label{RLE}
\end{figure}

Run Length Encoding (RLE) is the simplest lossless compression algorithm. The idea of RLE is to replace the sequences of the same data values by a single value followed by the number of occurrences as shown in Fig. \ref{RLE}. RLE is efficient with the data that contain lots of repetitive values \cite{akhter2010ecg,salomon2004data,drweesh2014audio}.
\subsection{Arithmetic Coding}
Arithmetic Coding is an entropy encoding algorithm used in lossless data compression. Arithmetic coding takes a message composed of symbols as input and converts it into a number (floating point number) less than one and greater than zero.

First, Arithmetic algorithm reads the input message (data file) symbol by symbol and starts with a certain interval. Then, it narrows the interval based on the probability of each symbol. Starting a new interval needs more bits. Therefore, arithmetic algorithm allows the low probability symbols to narrow the interval more than the high probability symbols and this is the key idea behind using the arithmetic encoding in the data compression field \cite{howard1991analysis,moffat1998arithmetic,witten1987arithmetic}.
\section{Proposed compression system for EEG data}
\begin{figure*}[h]
	\centering
	\includegraphics[width=\linewidth]{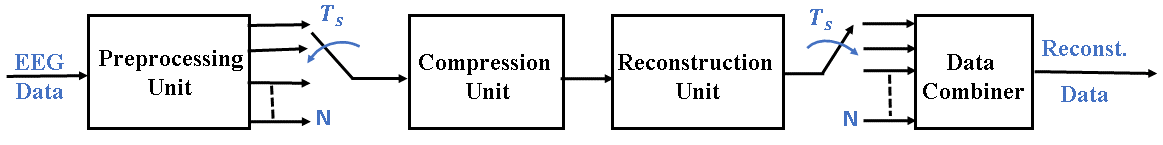}
	\captionof{figure}{Block diagram of the proposed compression system} \label{Blockdiagram_2}
\end{figure*}

The proposed compression system composed of a preprocessing unit, compression unit, reconstruction unit, and data combiner unit as shown in Fig. \ref{Blockdiagram_2}, . 
\subsection{Preprocessing Unit}
The functions of this unit are reading the recorded EEG data, standardizing, and segmenting it.

Standardization gives the EEG data the property of standard normal distribution which makes the EEG data in the same scale and this helps the compression unit to give high compression ratio. Standardization shifts the mean of the EEG data so that it is centered at zero with standard deviation of one as illustrated in Algorithm \ref{Preprocessing_Algorithm}. 

Let $X$ be a vector of the EEG data, the standardized EEG data $X_s$ is given as follows:
\begin{equation}
X_s=\frac{X-\mu}{\sigma}
\label{standardization}
\end{equation}
Where $\mu$ is the mean and $\sigma$ is the standard deviation of $X$.

The standardized EEG data are then segmented with sampling time $T_s$ in order to improve the compression performance. As shown in Fig. \ref{Blockdiagram_2}, the preprocessing unit produces one segment of the EEG data every $T_s$ second. Therefore, the size of the produced EEG segment depends on $T_s$. Decreasing $T_s$ improves the total compression time. However, the value of $T_s$ cannot be decreased lower than a threshold value in order to guarantee that each incoming EEG segment arrives a new unit after finishing the previous segment as follows:
\begin{equation}
T_{s}\geq max(T_{Lossy}, T_{thr}, T_{Lossless}, T_{Ilossless}, T_{Ilossy})
\label{Time_Condition}
\end{equation} 
Where $T_{Lossy}$ is the time of lossy compression algorithm, $T_{thr}$ is the thresholding time, $T_{Lossless}$ is the time of lossless compression algorithm, $T_{Ilossless}$ is the time of the inverse lossless algorithm, and $T_{Ilossy}$ is the time of the inverse lossy algorithm.

Therefore, the minimum sampling time ($T_s$) is obtained as follows: 
\begin{equation}
T_{min}=max(T_{Lossy}, T_{thr}, T_{Lossless}, T_{Ilossless}, T_{Ilossy})  
\end{equation}
The smallest compression and reconstruction time is achieved at $T_s=T_{min}$.
\begin{algorithm}[!ht]
	\caption{Preprocessing Algorithm}
	\label{Preprocessing_Algorithm}
	
	\hspace*{\algorithmicindent} \textbf{Input:} \text{Recorded EEG Data}\\
	\hspace*{\algorithmicindent} \textbf{Output:} \text{Preprocessed EEG Data}
	\smallskip
	\begin{algorithmic}[preprocessing]
		
		\LineComment {Standardization of EEG Data}
		\State $x \gets EEG data$
		\State $\mu \gets mean \; of \; x$
		\State $\sigma \gets standard \; deviation \; of \; x$
		\State $x_s=(x-\mu)/\sigma$
		
		\LineComment {Sampling}
		\State $N \gets NumberOfRequiredSamples$
		\State $L \gets LengthOfEEGData$
		\State $sp \gets floor(L/N)$
		\State $k \gets 1$
		\While{$k \leq N$}
		\If{$k=1$}
		\State $Data \gets EEGData(1:sp)$
		\Else 
		\State $initial \gets (k-1)*sp+1$
		\State $finals \gets k*sp$
		\State $Data \gets EEGData(initial:final)$
		\EndIf
		\If{$k=N$}
		\State $vector \gets EEGData(k*sp+1:L)$
		\State $Data \gets [Data\quad  vector]$
		\EndIf
		\State $k \gets k+sp$
		\EndWhile
	\end{algorithmic}
\end{algorithm}
\subsection{Compression Unit}
The compression unit composes of a lossy compression algorithm followed by a lossless compression algorithm. Both DCT and DWT are used as a lossy compression algorithms. Then, a thresholding is applied after the lossy compression algorithm in order to increase the redundancy of the transformed data. The values of the transformed data below a threshold value are set to zero. Therefore, varying the threshold value increases/decreases the number of zero coefficients. Consequently, the accuracy of the compression system is controllable based on the threshold value.

Finally, a lossless compression algorithm is applied. Both RLE and Arithmetic algorithms are proposed in this system. The lossless compression algorithms gives high compression ratio due to the high redundancy of the transformed data. 
\begin{figure*}[h]
	\centering
	\includegraphics[width=\linewidth]{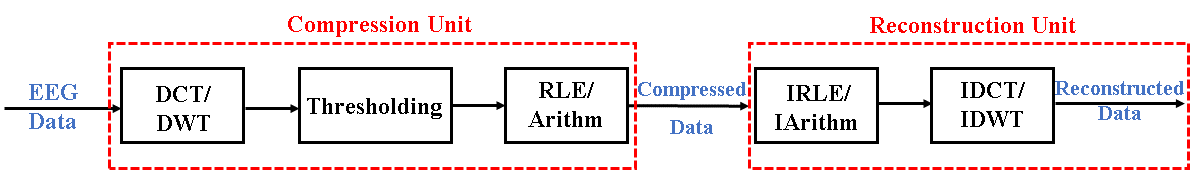}
	\captionof{figure}{Compression and Reconstruction Units} \label{Blockdiagram_1}
\end{figure*}
\subsection{Reconstruction Unit}
the inverse process of the compression unit is applied in this unit to reconstruct the original EEG data. First, the inverse of RLE/Arithmetic algorithm is applied. Then, the inverse of the DCT/DWT is applied to completely reconstruct the original EEG data as shown in Fig. \ref{Blockdiagram_1} and Algorithm \ref{Algorithm_2}.
\begin{algorithm}[!ht]
	\caption{Compression and Reconstruction Algorithm }
	\label{Algorithm_2}
	\hspace*{\algorithmicindent} \textbf{Input:} \text{Preprocessed EEG Data}\\
	\hspace*{\algorithmicindent} \textbf{Output:} \text{Recovered EEG Data}
	\smallskip
	\begin{algorithmic}[Compression]
		\LineComment {lossy Compression}
		\If{\textit{DCT is Selected}}
		\State $\textit{Transformed Data}\gets \text{DCT}\textit{(Preprocessed Data)}$
		\Else
		\State $\textit{Transformed Data}\gets \text{DWT}\textit{(Preprocessed Data)}$
		\EndIf
		\LineComment {Thresholding}
		\State $\textit{Thr}\gets \textit{Threshold Value}$
		\State $\textit{[Sorted Data, index]} \gets \textit{sort}(|\textit{Preprocessed Data}|)$ 
		\State $i \gets 1$
		\For{$\textit{Length of Data}$}
		\If{$|x(i)/x(1)|>Thr$}
		\State $i \gets i+1$
		\State {$continue$}
		\Else
		\State {$break$}
		\EndIf
		\EndFor
		\State $\textit{Transformed Data(index(i+1:end))} \gets 0$
		\LineComment {Lossless Compression}
		\If{$\textit{RLE is Required}$}
		\State $\textit{Compressed Data} \gets \text{RLE}\textit{(Transformed Data)}$
		\Else
		\State $\textit{Compressed Data}\gets \text{Arithm}\textit{(Transformed Data)}$
		\EndIf
		
		\\
		
		\LineComment {\textbf{Reconstruction Unit}}
		\If{$\textit{RLE is used}$}
		\State $\textit{Decoded Data}\gets \text{IRLE}\textit{(Compressed Datta)}$
		\Else
		\State $\textit{Decoded Data}\gets \text{IArithm}\textit{(Compressed Data)}$
		\EndIf
		\If{$\textit{DCT is used}$}
		\State $\textit{Reconstructed Data} \gets \text{IDCT}\textit{(Decoded Data)}$
		\Else
		\State $\textit{Reconstructed Data} \gets \text{IDWT}\textit{(Decoded Data)}$
		\EndIf	
		\LineComment {Data combining}
		\State $\textit{Final Output} \gets [\textit{Final Output} \quad \textit{Reconstructed Data}]$
	\end{algorithmic}
\end{algorithm}
\subsection{Performance Metrics}
The performance metrics used to evaluate the proposed compression system are described below:
\subsubsection{Root Mean Square Error (RMSE)}
The RMSE measures the error between two signals. Therefore, the RMSE is used in this paper to measure the the error between the original data and the reconstructed data. The RMSE is given by:
\begin{equation}
RMSE=\sqrt{\frac{\sum_{i=1}^{N}(X_{i}-Y_{i})^{2}}{N}}
\label{RMSE}
\end{equation} 
Where $X$ and $Y$ are the original and recovered data respectively.
\subsubsection{Compression Ratio (CR)}
The CR calculated as the difference in size between the original data and the compressed data divided by the size of the original data as follows:
\begin{equation}
CR=\frac{OriginalDataSize-CompDataSize}{OriginalDataSize}\times100 
\end{equation} 
\subsubsection{Compression and Reconstruction Time (T)}
The last performance metric used in this paper is the compression and reconstruction time which is given as follows:
\begin{equation}
T=T_{comp}+T_{reconst}
\end{equation}
where $T_{comp}$ and $T_{reconst}$ are the compression and recovering time respectively and defined as the following:
\begin{equation}
T_{comp}=T_{Lossy}+T_{thr}+T_{Lossless}
\end{equation}
\begin{equation}
T_{reconst}=T_{Ilossless}+T_{Ilossy}
\end{equation}
Finally, the total time is given as follows:
\begin{equation}
T=T_{Lossy}+T_{thr}+T_{Lossless}+T_{Ilossy}+T_{Ilossless}
\end{equation}
\section{Performance Evaluation}
The performance of the proposed compression system is evaluated using Python run on Intel(R) Core(TM) i3 3.9GHz CPU and 8GB RAM. The size of the used EEG data is 1 MB.

\begin{figure}[!ht]
	\centering
	\captionsetup{justification=centering}
	\includegraphics[width=\linewidth]{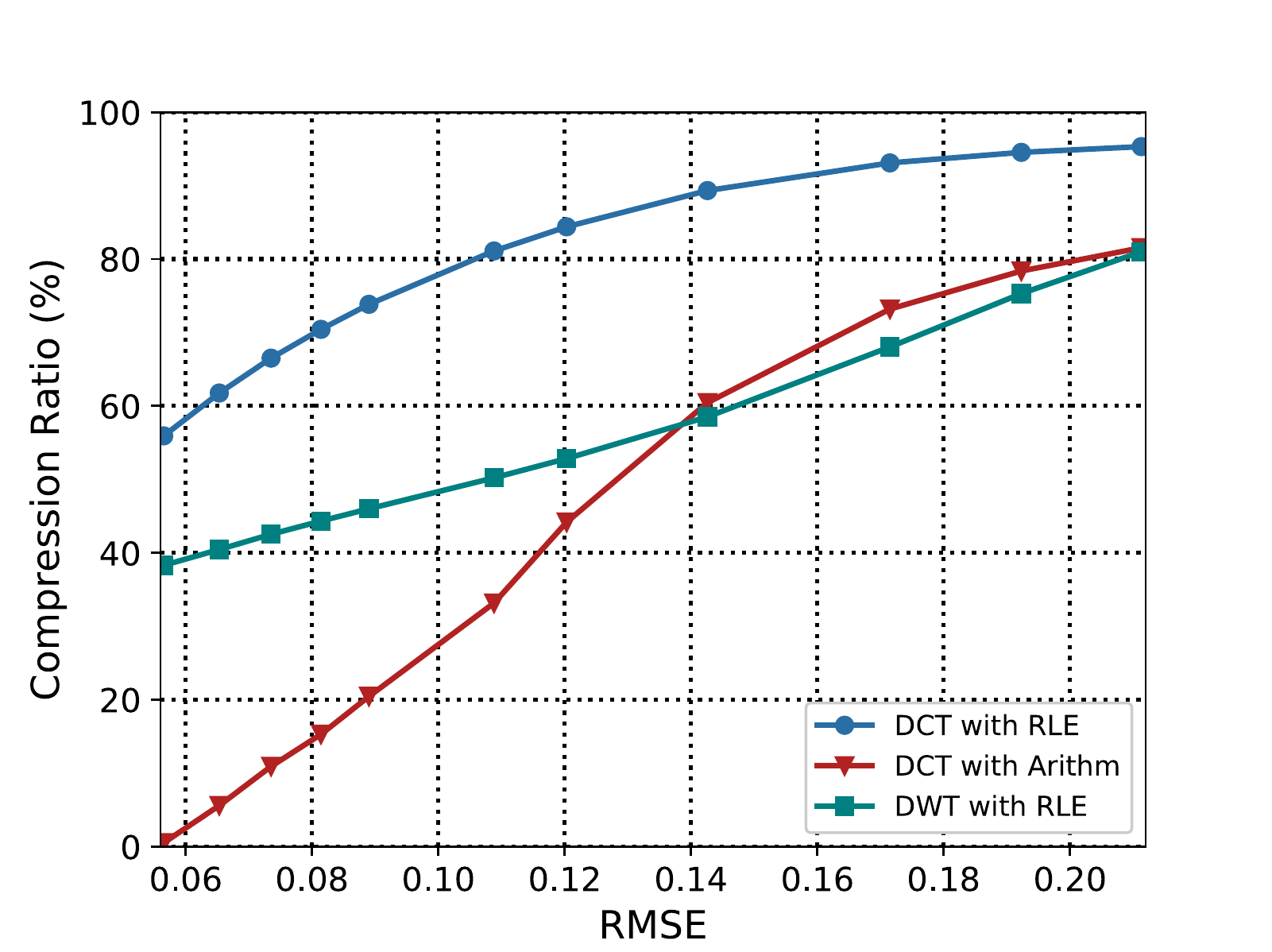}
	\caption{CR versus RMSE}
	\label{CR versus RMSE}
\end{figure}

\begin{figure}[!ht]
	\centering
	\captionsetup{justification=centering}
	\includegraphics[width=\linewidth]{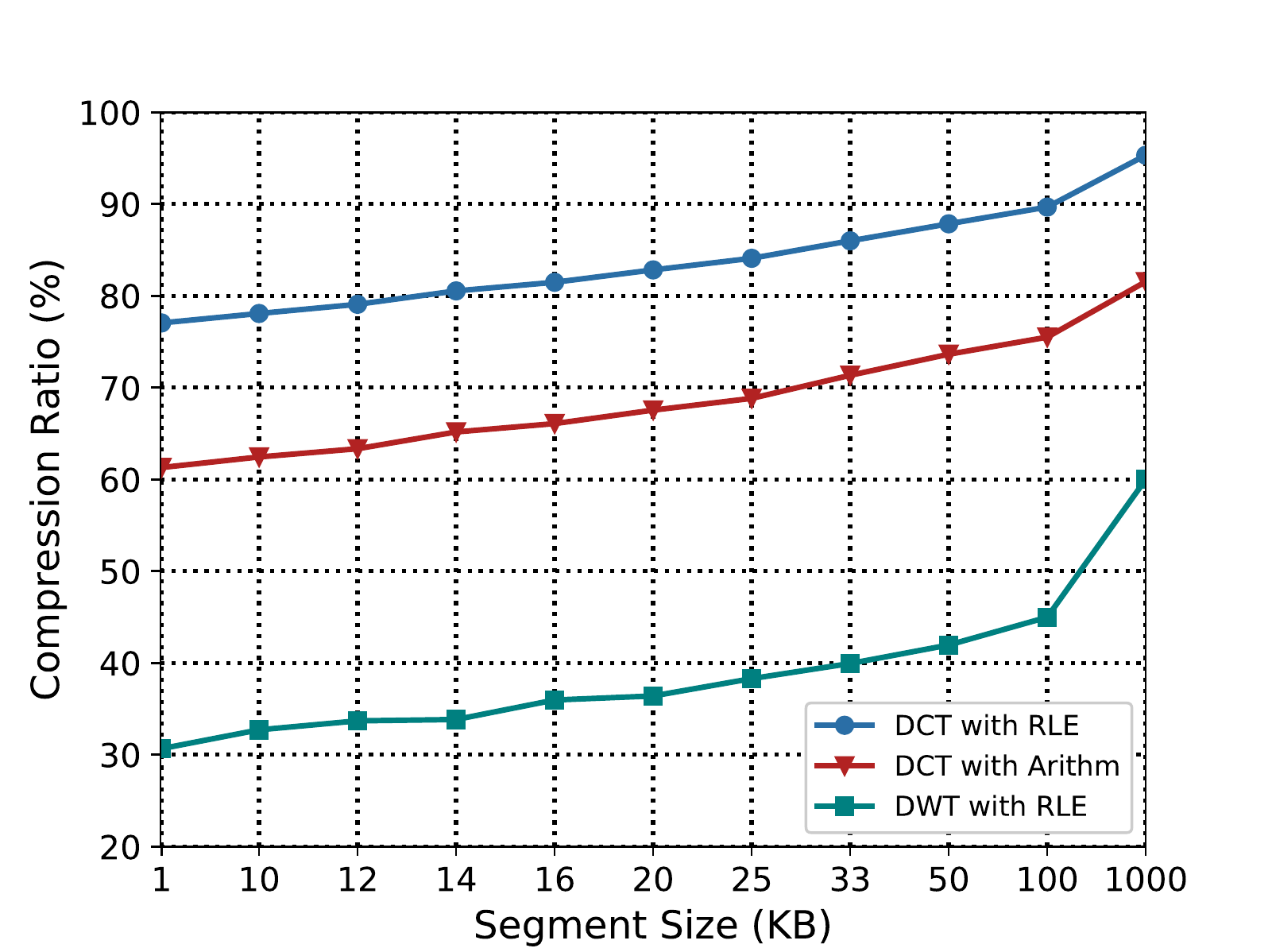}
	\caption{CR with segment size}
	\label{CR versus segment size}
\end{figure}

Fig. \ref{CR versus RMSE} shows the CR with different values of the RMSE. As shown in this figure, using the DCT as a lossy compression algorithm and RLE as a lossless compression technique gives the best CR compared with DCT/Arithm and DWT/RLE. The high CR of the DCT/RLE comes from the capability of the DCT to produce data with high redundancy and this facilitates the use of RLE. Both DCT/Arithm and DWT/RLE have approximately the same CR at high RMSE. The RMSE values are controlled by changing the threshold value. In these results, the threshold values are chosen between $0.005$ and $0.05$.  
Fig. \ref{CR versus segment size} shows the CR with segment size. As shown in this figure, the CR increases slightly if the segments size is increased. The size of the segment depends on the sampling time ($T_s$), \textit{i.e.,} increasing $T_s$ gives segments with large size and vice versa. Also, this figure shows that the DCT/RLE has the highest CR.

\begin{figure}[!ht]
	\centering
	\captionsetup{justification=centering}
	\includegraphics[width=\linewidth]{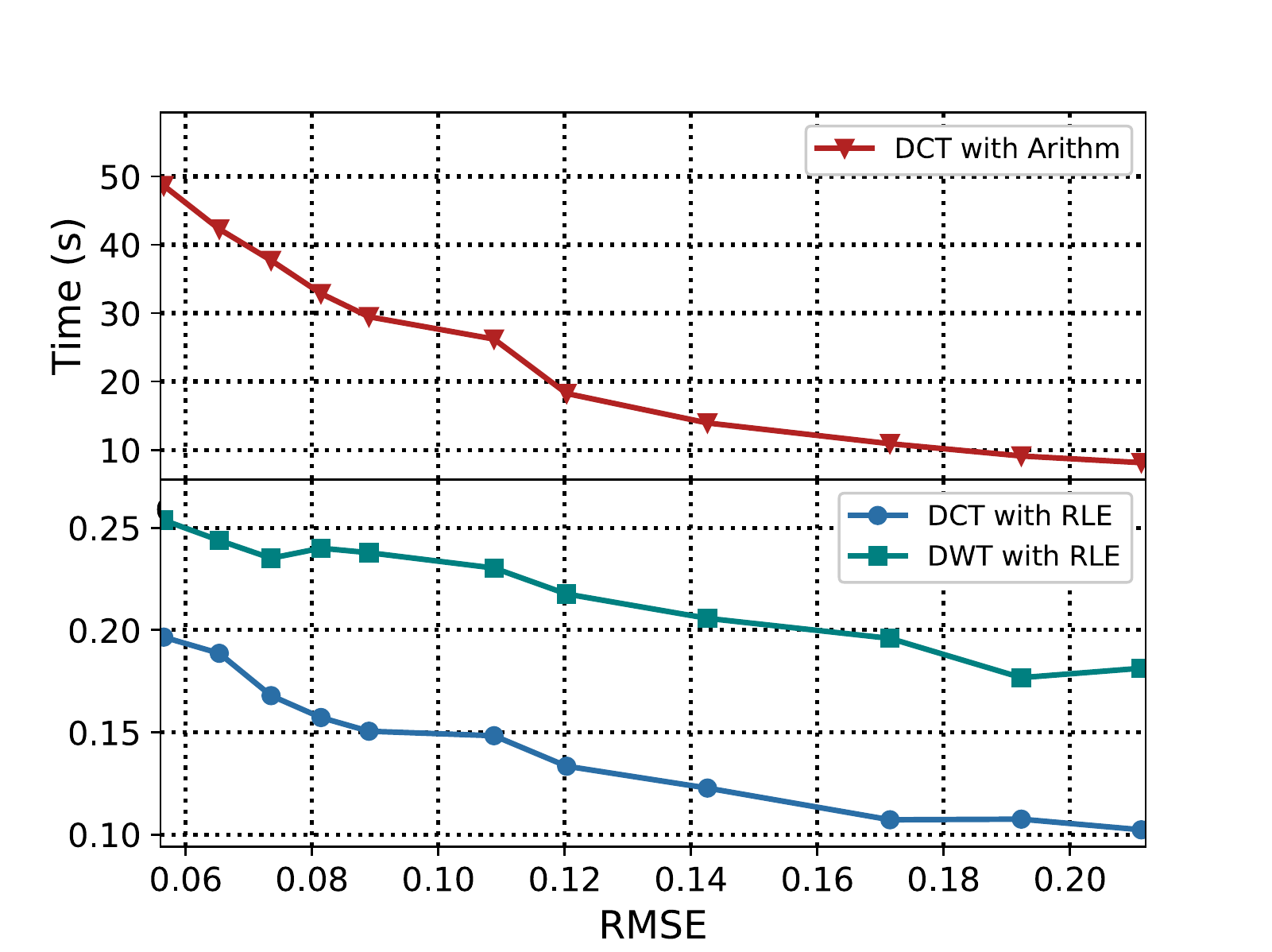}
	\caption{Compression and Reconstruction Time with RMSE}
	\label{Time_RMSE}
\end{figure}

\begin{figure}[!ht]
	\centering
	\captionsetup{justification=centering}
	\includegraphics[width=\linewidth]{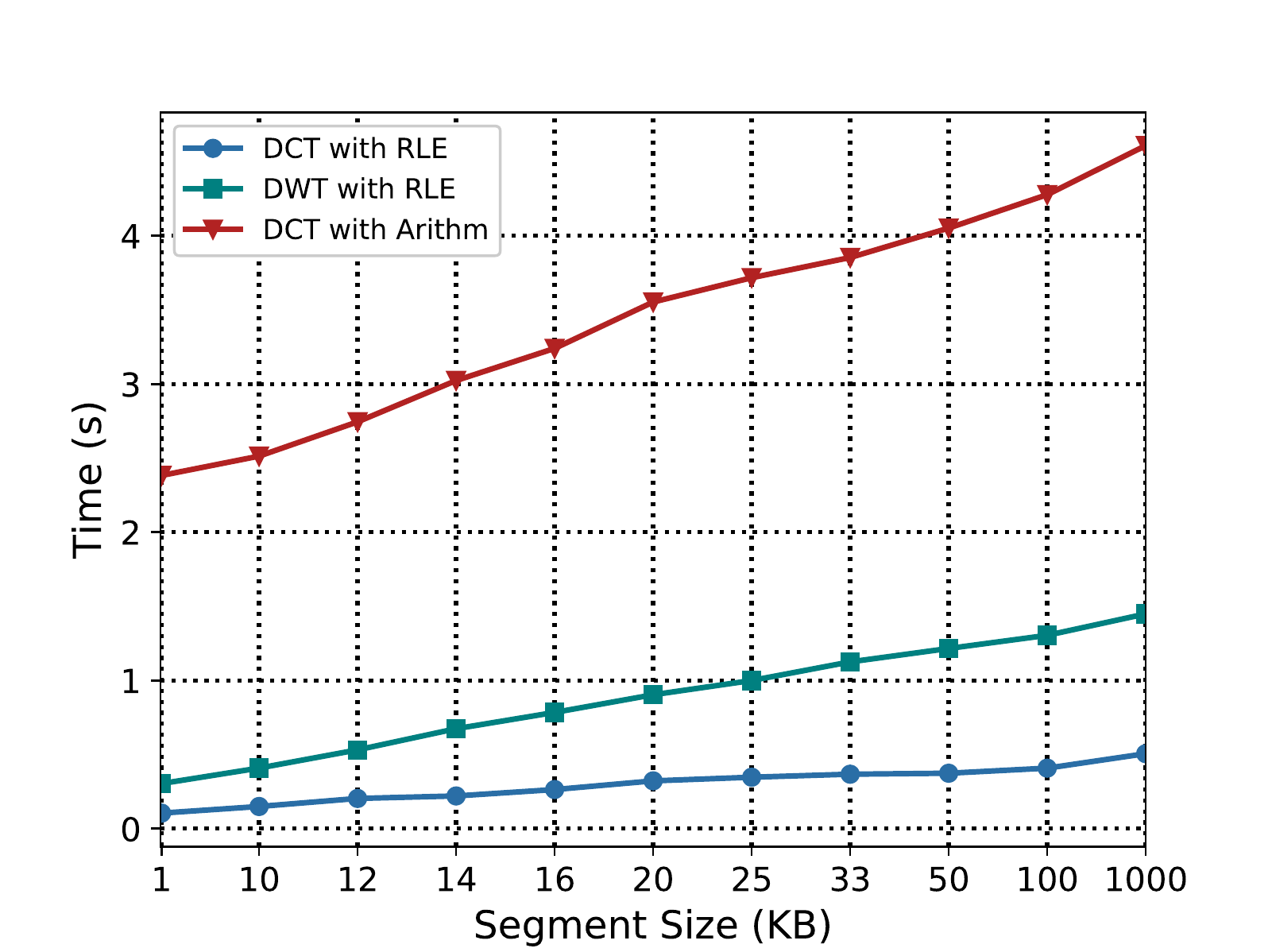}
	\caption{Compression and Reconstruction Time with Segment Size}
	\label{Time_segment size}
\end{figure}

The compression and reconstruction time ($T$) with RMSE of all studied cases is shown in Fig. \ref{Time_RMSE}. We can notice from this figure that both DCT/RLE and DWT/RLE take short time due to its simplicity, whereas DCT/Arithm is more complex and consumes longer time. 
Fig. \ref{Time_segment size} shows the compression and reconstruction time with segment size. as shown in this figure, increase the segment size causes increase in the compression time. Therefore, there is a trade off between CR and compression time when setting the sampling time ($T_s$) which controls the segment size.

\begin{figure}[!ht]
	\centering
	\captionsetup{justification=centering}
	\includegraphics[width=\linewidth]{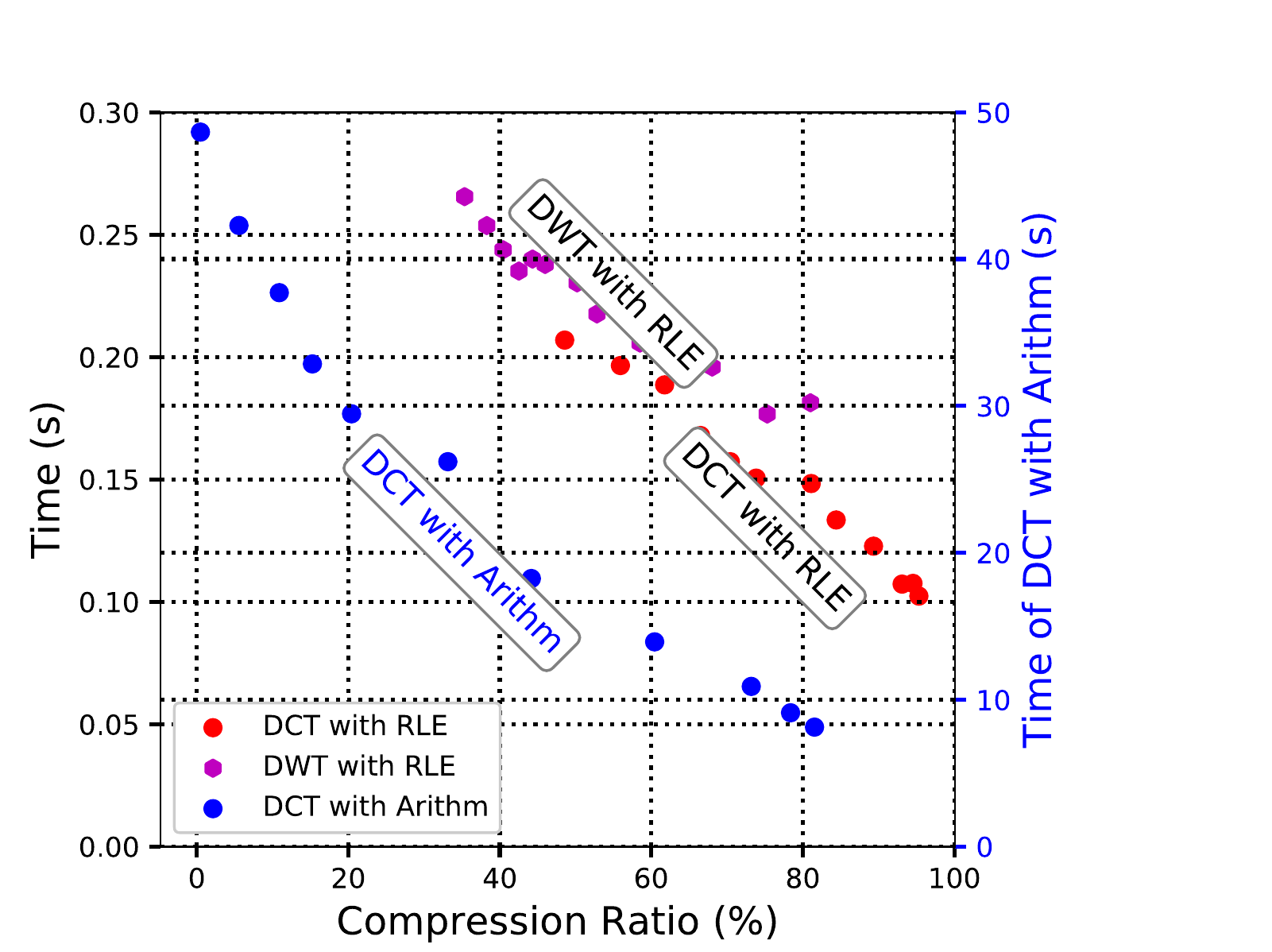}
	\caption{Compression and Reconstruction Time versus Compression Ratio with different values of RMSE}
	\label{final_comparison}
\end{figure}

The comparison between all the proposed compression algorithms in terms of CR and $T$ is shown in Fig. \ref{final_comparison}. As shown in this figure, DCT/RLE gives the best results in both CR and compression time. Also, DWT/RLE gives a good compression time and accepted CR whereas DCT/Arithm consumes long time compared with DCT/RLE and DWT/RLE. 
   
\begin{figure}[!ht]
	\centering
	\captionsetup{justification=centering}
	\includegraphics[width=0.5\textwidth]{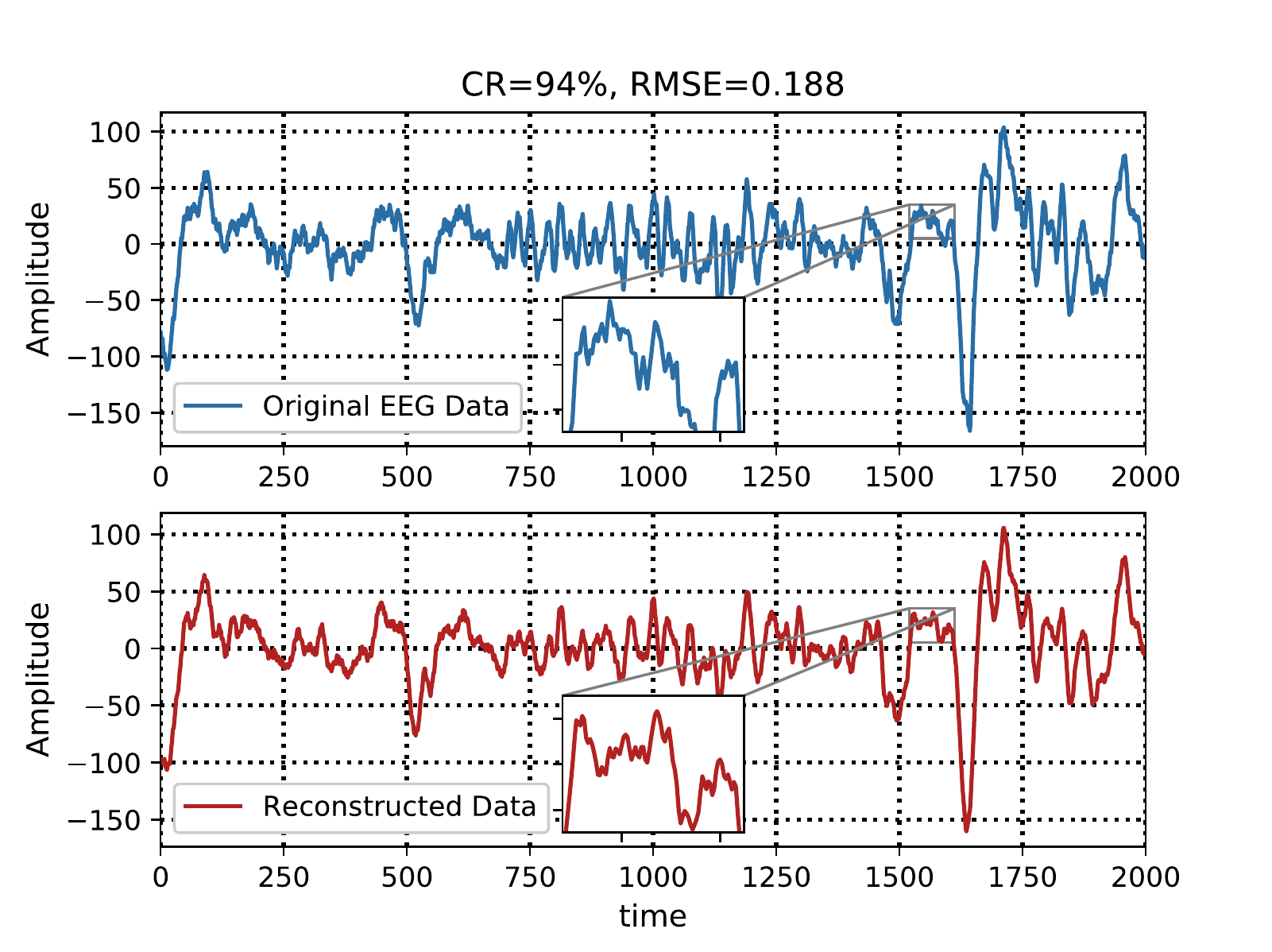}
	\includegraphics[width=0.5\textwidth]{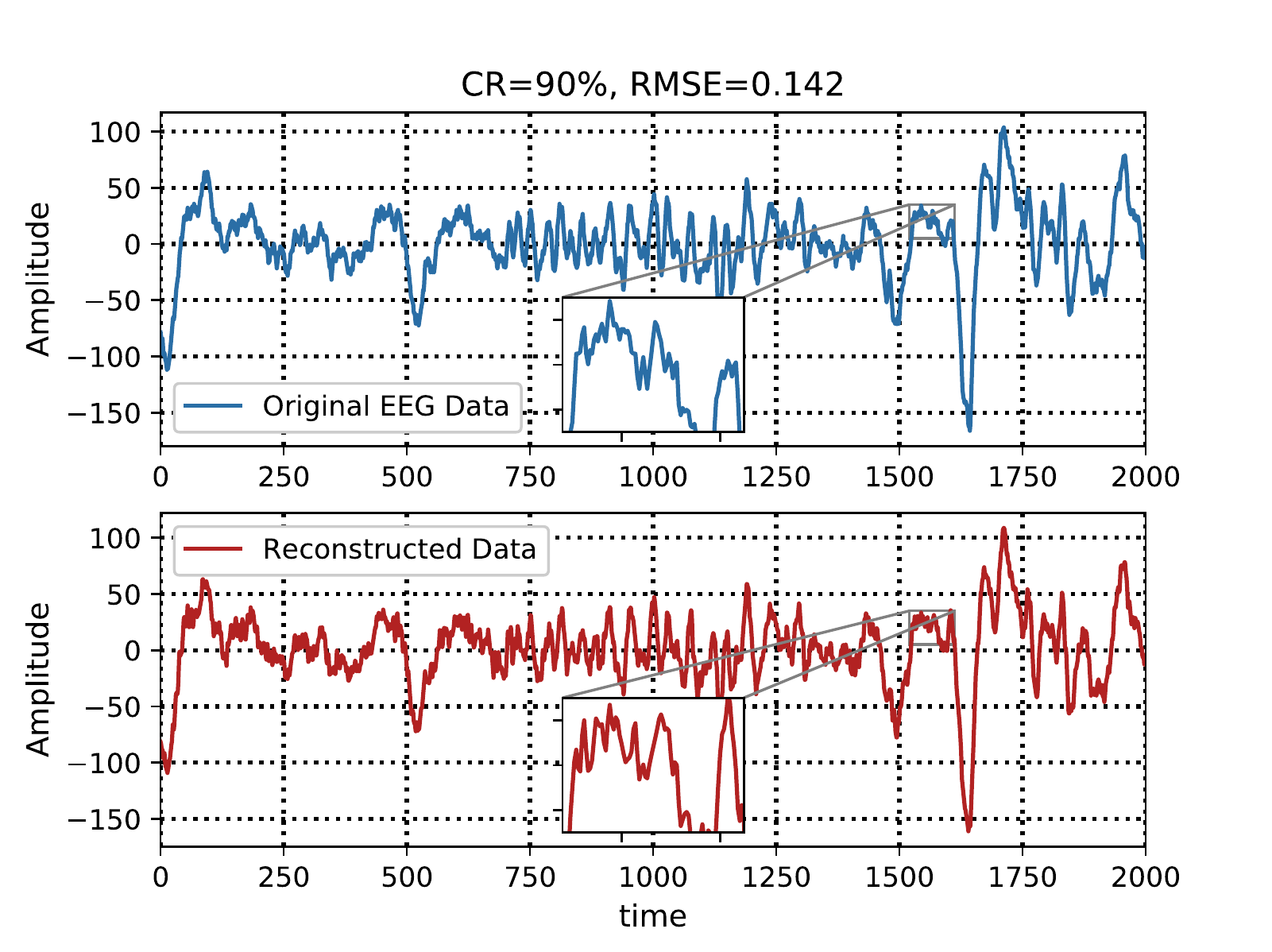}
	\includegraphics[width=0.5\textwidth]{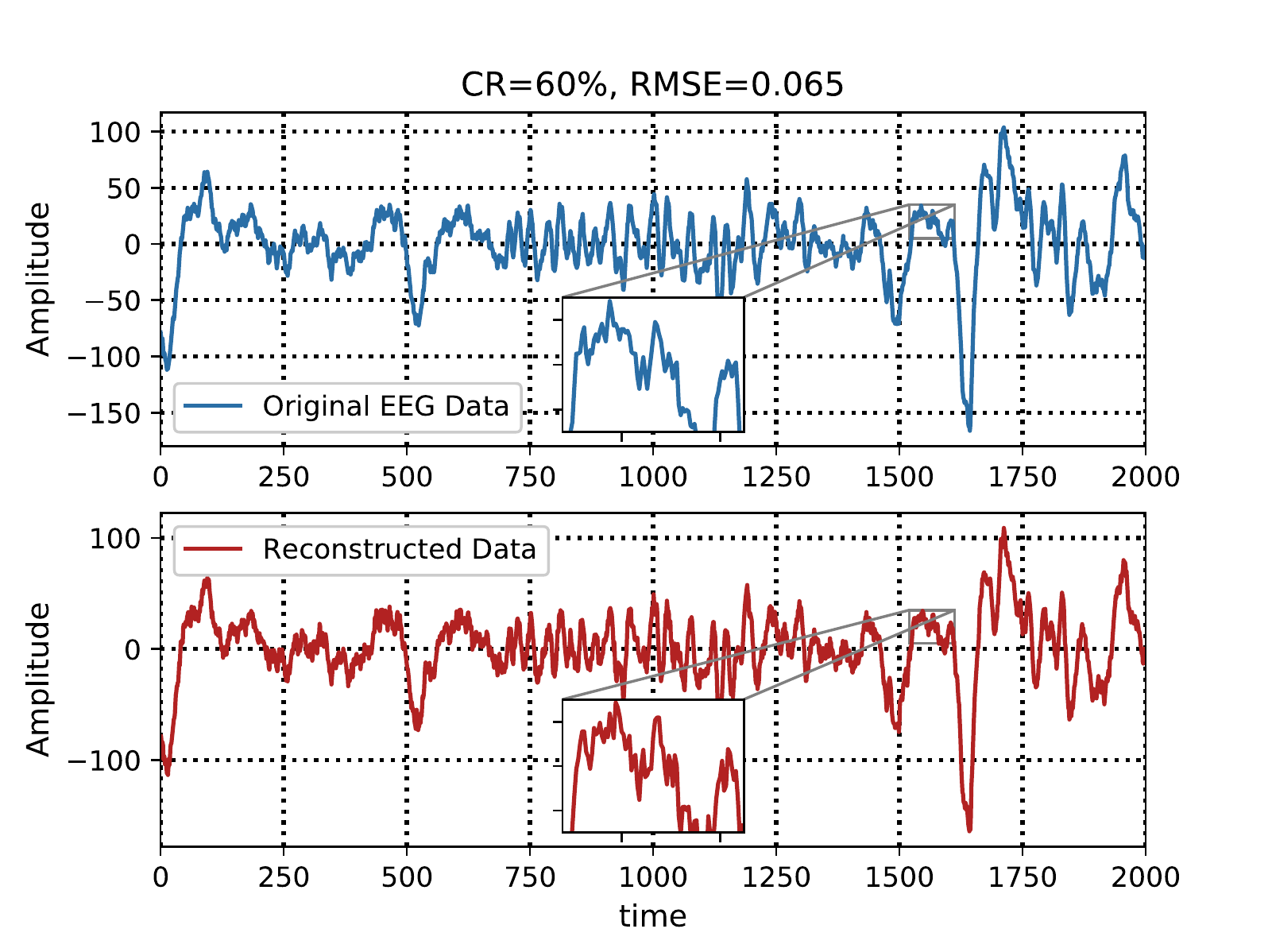}
	\caption{Original and Reconstructed EEG data with different CR in case of DCT/RLE }
	\label{Original_vs_Reconst_EEG_data}
\end{figure}

Finally, Fig. \ref{Original_vs_Reconst_EEG_data} shows the original and the reconstructed EEG data with different CR in case of DCT with RLE. As shown in Fig. \ref{Original_vs_Reconst_EEG_data}, there is small distortion in the recovered EEG data at $CR=95\%$, \textit{i.e.,} the $RMSE=0.188$, whereas at $CR=60$ the distortion is less and both the original and recovered EEG data approximately the same, \textit{i.e.,} the $RMSE=0.065$.
\section{Conclusion}
A compression system composed of both lossy and lossless compression algorithms is designed in this article. The DCT and the DWT transforms followed by thresholding are used as a lossy compression technique. We have used the RLE and Arithmetic encoding as a lossless compression algorithms. The data produced by the lossy compression part contains high redundancy and this facilitates the use of lossless algorithms. CR, RMSE and compression time are calculated in order to check the performance of the system. We conclude that using DCT as a lossy compression algorithm followed by RLE as a lossless compression algorithm gives the best performance compared with DWT and Arithmetic encoding. As a future work, we will implement the technique of DCT with RLE on hardware for EEG data compression and check its performance in real implementation.    
\section*{Acknowledgement}
This work was supported by the Egyptian Information Technology Industry Development Agency (ITIDA) under ITAC Program CFP 96.
\bibliographystyle{IEEEtran}
\bibliography{Ref}

\begin{thebibliography}{10}
\providecommand{\url}[1]{#1}
\csname url@samestyle\endcsname
\providecommand{\newblock}{\relax}
\providecommand{\bibinfo}[2]{#2}
\providecommand{\BIBentrySTDinterwordspacing}{\spaceskip=0pt\relax}
\providecommand{\BIBentryALTinterwordstretchfactor}{4}
\providecommand{\BIBentryALTinterwordspacing}{\spaceskip=\fontdimen2\font plus
\BIBentryALTinterwordstretchfactor\fontdimen3\font minus
  \fontdimen4\font\relax}
\providecommand{\BIBforeignlanguage}[2]{{%
\expandafter\ifx\csname l@#1\endcsname\relax
\typeout{** WARNING: IEEEtran.bst: No hyphenation pattern has been}%
\typeout{** loaded for the language `#1'. Using the pattern for}%
\typeout{** the default language instead.}%
\else
\language=\csname l@#1\endcsname
\fi
#2}}
\providecommand{\BIBdecl}{\relax}
\BIBdecl

\bibitem{Alsenwi2017}
M.~Alsenwi, M.~Saeed, T.~Ismail, H.~Mostafa, and S.~Gabran, ``Hybrid
  compression technique with data segmentation for electroencephalography
  data,'' in \emph{2017 29th International Conference on Microelectronics
  ({ICM})}.\hskip 1em plus 0.5em minus 0.4em\relax {IEEE}, dec 2017.

\bibitem{birvinskas2015fast}
D.~Birvinskas, I.~Jusas, and Damasevicius, ``Fast dct algorithms for eeg data
  compression in embedded systems,'' \emph{Computer Science and Systems},
  vol.~12, no.~1, pp. 49--62, 2015.

\bibitem{Alsenwi2016}
M.~Alsenwi, T.~Ismail, and H.~Mostafa, ``Performance analysis of hybrid
  lossy/lossless compression techniques for {EEG} data,'' in \emph{2016 28th
  International Conference on Microelectronics ({ICM})}.\hskip 1em plus 0.5em
  minus 0.4em\relax {IEEE}, dec 2016.

\bibitem{hadjileontiadis2006biosignals}
L.~J. Hadjileontiadis, ``Biosignals and compression standards,'' in
  \emph{M-Health}.\hskip 1em plus 0.5em minus 0.4em\relax Springer, 2006, pp.
  277--292.

\bibitem{akhter2010ecg}
S.~Akhter and M.~Haque, ``Ecg comptression using run length encoding,'' in
  \emph{Signal Processing Conference, 2010 18th European}.\hskip 1em plus 0.5em
  minus 0.4em\relax IEEE, 2010, pp. 1645--1649.

\bibitem{deshlahra2013comparative}
A.~Deshlahra, G.~Shirnewar, and A.~Sahoo, ``A comparative study of dct, dwt \&
  hybrid (dct-dwt) transform,'' 2013.

\bibitem{antoniol1997eeg}
G.~Antoniol and P.~Tonella, ``Eeg data compression techniques,''
  \emph{Biomedical Engineering, IEEE Transactions on}, vol.~44, no.~2, pp.
  105--114, 1997.

\bibitem{koyrakh2008data}
L.~Koyrakh, ``Data compression for implantable medical devices,'' in
  \emph{Computers in Cardiology, 2008}.\hskip 1em plus 0.5em minus 0.4em\relax
  IEEE, 2008, pp. 417--420.

\bibitem{salomon2004data}
D.~Salomon, \emph{Data compression: the complete reference}.\hskip 1em plus
  0.5em minus 0.4em\relax Springer Science and Business Media, 2004.

\bibitem{fauvel2014energy}
S.~Fauvel and Ward, ``An energy efficient compressed sensing framework for the
  compression of electroencephalogram signals,'' \emph{Sensors}, vol.~14,
  no.~1, pp. 1474--1496, 2014.

\bibitem{rajoub2002efficient}
B.~A. Rajoub, ``An efficient coding algorithm for the compression of ecg
  signals using the wavelet transform,'' \emph{Biomedical Engineering, IEEE
  Transactions on}, vol.~49, no.~4, pp. 355--362, 2002.

\bibitem{shaeri2015method}
M.~A. Shaeri and A.~M. Sodagar, ``A method for compression of
  intra-cortically-recorded neural signals dedicated to implantable
  brain--machine interfaces,'' \emph{Neural Systems and Rehabilitation
  Engineering, IEEE Transactions on}, vol.~23, no.~3, pp. 485--497, 2015.

\bibitem{khalifa2008compression}
O.~O. Khalifa, S.~H. Harding, and A.-H. Abdalla~Hashim, ``Compression using
  wavelet transform,'' \emph{International Journal of Signal Processing},
  vol.~2, no.~5, pp. 17--26, 2008.

\bibitem{drweesh2014audio}
Z.~T. DRWEESH and L.~E. GEORGE, ``Audio compression based on discrete cosine
  transform, run length and high order shift encoding,'' \emph{International
  Journal of Engineering and Technology (IJEIT (, Vol. 4, Issue 1, Pp. 45-51},
  2014.

\bibitem{howard1991analysis}
P.~G. Howard and J.~S. Vitter, ``Analysis of arithmetic coding for data
  compression,'' in \emph{Data Compression Conference, 1991. DCC'91.}\hskip 1em
  plus 0.5em minus 0.4em\relax IEEE, 1991, pp. 3--12.

\bibitem{moffat1998arithmetic}
A.~Moffat, R.~M. Neal, and Witten, ``Arithmetic coding revisited,''
  \emph{Transactions on Information Systems (TOIS)}, vol.~16, no.~3, pp.
  256--294, 1998.

\bibitem{witten1987arithmetic}
I.~H. Witten, R.~M. Neal, and Cleary, ``Arithmetic coding for data
  compression,'' \emph{Communications of the ACM}, vol.~30, no.~6, pp.
  520--540, 1987.

\end{thebibliography}
\end{document}